\documentclass[%
reprint,
showpacs,preprintnumbers,
nofootinbib,
nobibnotes,
amsmath,amssymb,
prb,
floatfix,
longbibliography
]{revtex4-2}

\usepackage{graphicx}
\usepackage{dcolumn}
\usepackage{bm}
\usepackage{colortbl}
\usepackage[unicode=true,colorlinks=true,citecolor=blue,urlcolor=blue]{hyperref}
\usepackage{braket}
\usepackage[normalem]{ulem}
\usepackage{enumitem}

\newcommand{\sign}{\mathop{\rm sign}}
\newcommand{\e}{\mathrm{e}}

\let\ifr\i
\renewcommand{\i}{{\rm i}}
\renewcommand{\d}{\mathrm d}

\usepackage{chngcntr}
\newcommand{\enquote}{}
\newcommand{\nix}[1]{}
\let\oldsec\section
\renewcommand{\section}[1]{\textit{#1}---}

\begin{document}

\title{Valley-controlled chiral magnetism in transition metal dichalcogenide monolayers}

\author{I.~S.~Krivenko}
\affiliation{University of Hamburg, I. Institute of Theoretical Physics, 22607 Hamburg, Germany}
\author{V.~N.~Mantsevich}
\affiliation{Lomonosov Moscow State University, 119991 Moscow, Russia}
\author{D.~S. Smirnov}
\affiliation{Ioffe Institute, 194021 St.~Petersburg, Russia}
\email[Electronic address: ]{smirnov@mail.ioffe.ru}

\begin{abstract}
  We put forward the Ruderman--Kittel--Kasuya--Yosida (RKKY) interaction in transition metal dichalcogenide monolayers as a tool to create and control a chiral magnetic texture. We show that in the spin-valley locking regime, the RKKY interaction acts as a Dzyaloshinskii-Moriya coupling with an effective spin rotation period exactly equal to the tripled lattice constant. Using mean field theory and classical Monte Carlo simulations, we demonstrate that this interaction qualitatively reshapes the phase diagram of atomically thin antiferromagnets. It destroys the chirality-related phase transition by selecting a single chirality value even when the RKKY interaction is small. At the same time, it shifts the Berezinskii--Kosterlitz--Thouless transition associated with spins orientation to higher temperatures. We argue that the valley degree of freedom of electrons mediating the RKKY interaction provides a powerful control knob for exploring non-universal phase transitions and quantum spin liquid states in two-dimensional van der Waals heterostructures.
\end{abstract}

\maketitle{}

\textit{Introduction.---}The discovery of intrinsic two-dimensional (2D) magnetism has opened a new frontier in condensed matter physics, enabling the exploration of magnetic order in the ultimate thickness limit and fostering novel spintronic functionalities~\cite{doi:10.1126/science.aar4851,wang_NatureCommunications2018,wang_NatureNanotechnology2018,Huang2018,Gong2019}. Following the first demonstrations of long-range magnetic order in atomically thin crystals such as Cr$_2$Ge$_2$Te$_6$~\cite{Gong2017,deng_Nature2018a} and CrI$_3$~\cite{huang_Nature2017a,wang_NatureNanotechnology2019}, an expanding family of 2D magnets has revealed a rich spectrum of phenomena absent in their bulk counterparts~\cite{Fei2018,chua_AdvancedMaterials2021,son_ACSNano2021,lee_NanoLett2021,telford_NatureMaterials2022,long_NanoLett2020,ni_NatureNanotechnology2021,aapro_ACSNano2021,su_AdvancedScience2020,lin_NanoLett2021,chua_AdvancedMaterials2020}. 
However, achieving efficient and versatile control of magnetism in these materials remains a central challenge.

A promising route here is to couple 2D magnets to semiconducting van der Waals layers, whose carrier density and Fermi level can be tuned over a broad range by doping or electrostatic gating~\cite{Burch2018,Gibertini2019,sierra_NatureNanotechnology2021}. Across atomically sharp interfaces, proximity effects can transfer spin, valley, and magnetic correlations between different constituents, allowing one layer to act as an active medium for controlling another~\cite{macneill_NaturePhysics2017,Seyler2018,shi_NatureNanotechnology2019,Ciorciaro2020,wu_NatureCommunications2020,liang_AdvancedMaterials2020}. Transition metal dichalcogenide monolayers (TMD MLs) are especially attractive in this context: beyond their electrostatic tunability, they possess strong spin--orbit coupling and valley-selective physics~\cite{cao_NatureCommunications2012,xiao_PhysRevLett2012,MX2Review}. These properties make TMD MLs promising active components for introducing electronically controllable degrees of freedom into low-dimensional magnetism. In particular, their coupled spin and valley degrees of freedom suggest a route to magnetic exchange interactions that can be governed by the valley structure of itinerant carriers.

Here, we propose and theoretically demonstrate a mechanism for controlling magnetic order via the valley degree of freedom in atomically thin hybrid systems. We show that conduction electrons in TMD MLs mediate a chiral RKKY interaction~\cite{Ruderman1954,Kasuya1956,Yosida1957} between localized magnetic moments, enabling valley-dependent long-range exchange. This interaction provides a platform to explore the interplay between the Mermin--Wagner theorem~\cite{Mermin1966} and topological phase transitions~\cite{Berezinskii1971,Kosterlitz1973}. We find that its competition with direct antiferromagnetic exchange between magnetic impurities fundamentally reshapes the magnetic phase diagram: it suppresses the chirality-driven second order phase transition, selects a unique chirality determined by the sign of the conduction-band spin splitting, and shifts the Berezinskii--Kosterlitz--Thouless transition temperature. These results identify valley-dependent exchange as a versatile control knob for chiral magnetism in two dimensions, providing a route toward programmable magnetic phases and valley-enabled spintronic and quantum devices.

\begin{figure}[h]
  \centering
  \includegraphics[width=0.9\linewidth]{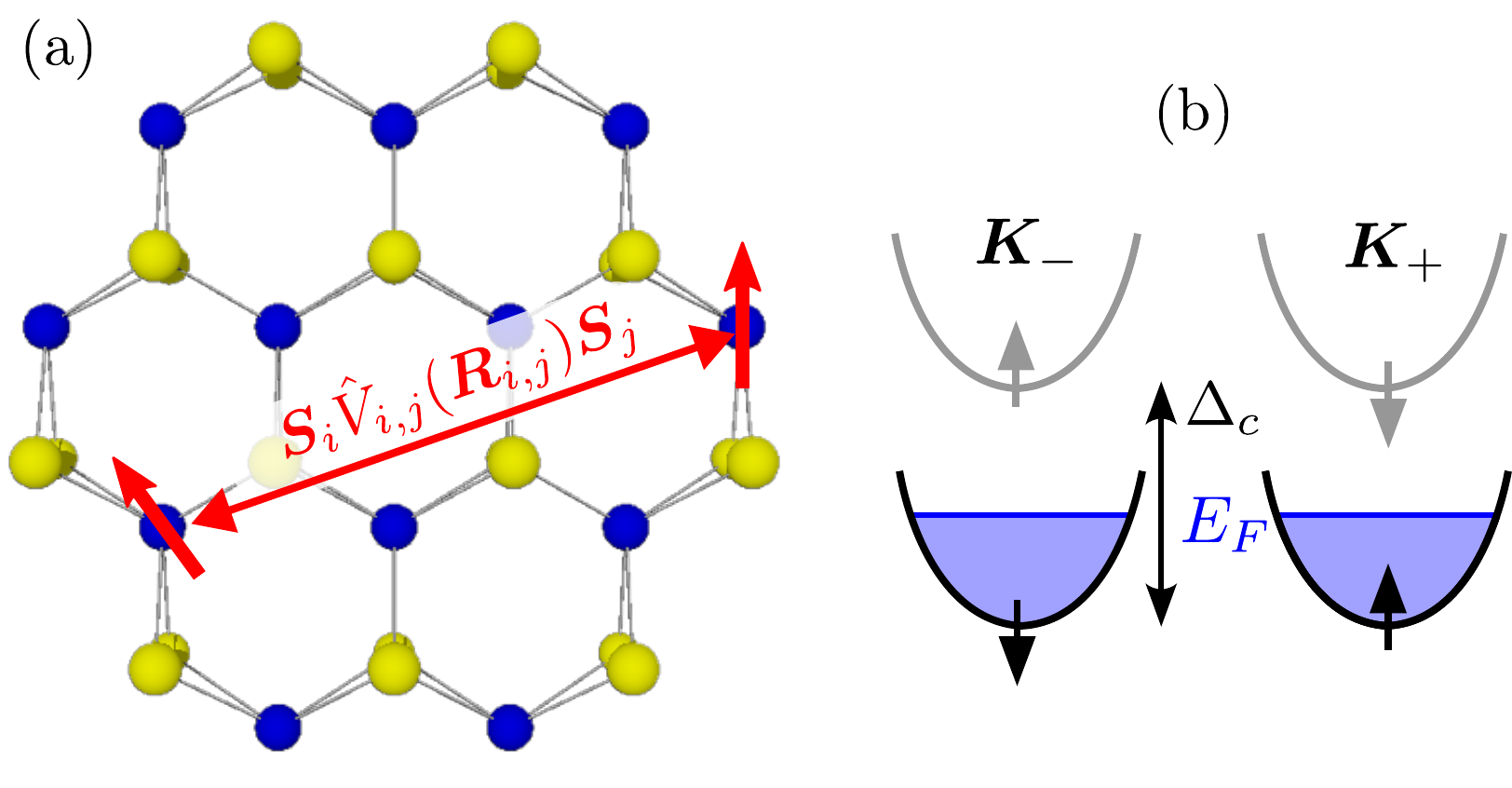}%
  \caption{(a) RKKY exchange interaction between two magnetic impurities (red arrows) on top of a TMD ML (yellow and blue spheres). (b) Spin-valley locking regime for the conduction electrons in the TMD ML: the electrons occupy the two lower spin subbands in the $K_+$ and $K_-$ valleys of the conduction band.} 
  \label{fig:scheme}
\end{figure}

\textit{Chiral RKKY interaction.---}As a first step toward 2D chiral magnetism, we consider the exchange interaction between two magnetic adatoms on top of a TMD ML, see Fig.~\ref{fig:scheme}(a). The chirality of the interaction arises due to the valley degeneracy of the conduction band with the minima at the $\bm K_\pm=(\pm\frac{4\pi}{3a_0},0)$ points of the Brillouin zone with $a_0$ being the lattice constant. We assume that the RKKY interaction is mediated by the resident electrons, which can be introduced to the structure by doping or application of a gate voltage. TMD MLs intrinsically lack inversion symmetry, so the two valleys experience spin splitting $\Delta_c$, typically on the order of 10~meV~\cite{kormanyos_PhysRevB2013,DurnevUFN,robert_NatCommun2020}, as shown in Fig.~\ref{fig:scheme}(b). We consider the Fermi level $E_F$ to lie below the upper spin-split subband, which is known as the spin-valley locking regime: In this case the electron spin is strictly locked to the valley index, see Fig.~\ref{fig:scheme}(b). 

The Hamiltonian of the exchange interaction between magnetic impurities and 2D electrons reads
\begin{equation}
  \mathcal H_0=\sum_{i}\bm S_i\hat{A}\bm\sigma(\bm R_i),
\end{equation}
where $\bm S_i$ are the spins of magnetic impurities at positions $\bm R_i$, $\hat{A}$ is the exchange interaction tensor, and $\bm\sigma(r)=\psi_s^\dag(\bm r)\bm\sigma_{s,s'}\psi_{s'}(\bm r)$ is the spin polarization of the electron gas described by the spinor operators $\psi_s(\bm r)$ (we implicitly assume summation over the spin indices $s,s'=\pm$). In the second order of the perturbation theory in the exchange interaction, we obtain in a standard way~\cite{giuliani2008quantum} the energy of RKKY interaction between impurities
\begin{equation}
  \label{eq:A2}
  \mathcal H_\text{ind}=-\sum_{i<j}\bm S_i\hat{A}\hat{\chi}_0(\bm R_{i,j})\hat{A}^T\bm S_j
\end{equation}
with the static spin susceptibility of the 2D electron gas 
\begin{equation}
  \label{eq:chi_main}
  \chi_0^{\alpha\beta}(\bm r)=\i\hbar\int\frac{\d\omega}{2\pi}\sigma_{ss'}^\alpha G_{s'}^{(0)}(\omega,\bm r)\sigma_{s's}^\beta G_{s}^{(0)}(\omega,-\bm r).
\end{equation}
Here, the components of the susceptibility tensor $\chi_0^{\alpha\beta}$ ($\alpha,\beta=x,y,z$) are expressed through the bare Green's function of the electrons in the frequency-coordinate representation $G_{s}^{(0)}(\omega,\bm r)$.

Due to the spin valley locking, the Green's function differs from the standard Green's function of the 2D electrons by a spin-dependent factor $G_{s}^{(0)}\propto\e^{\i\bm{K}_s\bm{r}}$. In the End Matter Appendix \hyperref[RKKYDerivation]{A} we show that this leads to the final expression for the indirect RKKY interaction
\begin{equation}
  \label{eq:H_RKKY}
  \mathcal H_\text{ind}=\sum_{i<j}\bm S_i\hat{V}(\bm{R}_{i,j})\bm S_j
\end{equation}
with $\bm R_{i,j}=\bm R_i-\bm R_j$ and the pairwise exchange interaction energy tensor
\begin{multline}
  \label{eq:V_tensor}
  \hat{V}(\bm R)=
  \begin{pmatrix}
    V_\parallel\cos\varphi & V_\parallel\sin\varphi & 0 \\
    -V_\parallel\sin\varphi & V_\parallel\cos\varphi & 0 \\
    0 & 0 & V_\perp
  \end{pmatrix}\\\times\left[J_0(k_FR)Y_0(k_FR)+J_1(k_FR)Y_1(k_FR)\right].
\end{multline}
Here $V_{\parallel,\perp}=m k_F^2A_{\parallel,\perp}^2/(2\pi\hbar^2)$ are the exchange interaction constants for the spin components parallel and perpendicular to the ML, respectively, with the corresponding components of the tensor $\hat{A}$ squared, $m$ being the electron effective mass and $k_F$ being the Fermi wave vector ($\hbar^2k_F^2/(2m)=E_F$), $\varphi=2|\bm K_\pm|R_x$, $J_{0,1}(x)$ and $Y_{0,1}(x)$ are Bessel functions of the first and second kinds, respectively.

Crucially, the phase $\varphi$ in Eq.~\eqref{eq:V_tensor} leads to the chirality of the exchange interaction, i.e. effective relative in-plane rotation of the interacting spins. Physically, it is related to the intervalley scattering of electrons by short range potential of magnetic impurities~\cite{Cullen1968} and to the corresponding spin-valley locking of the electrons. The form of this interaction is similar to the Dzyaloshinskii-Moriya interaction~\cite{Fang2021} but with a very short period of $3a_0$\footnote{Mathematically, the period equals $3a_0/4$, but physically only a multiple of the lattice constant makes sense.}. It requires broken inversion symmetry, which is intrinsic to TMD MLs and manifests itself in the spin splitting of the bands. For the opposite sign of the conduction band spin splitting, the sign of $\varphi$ should be reversed. Note that the chiral exchange interaction appears in this system without need for strain or twist~\cite{Cui2020,Osterholt2026}.

\begin{figure}[h]
  \centering
  \includegraphics[width=\linewidth]{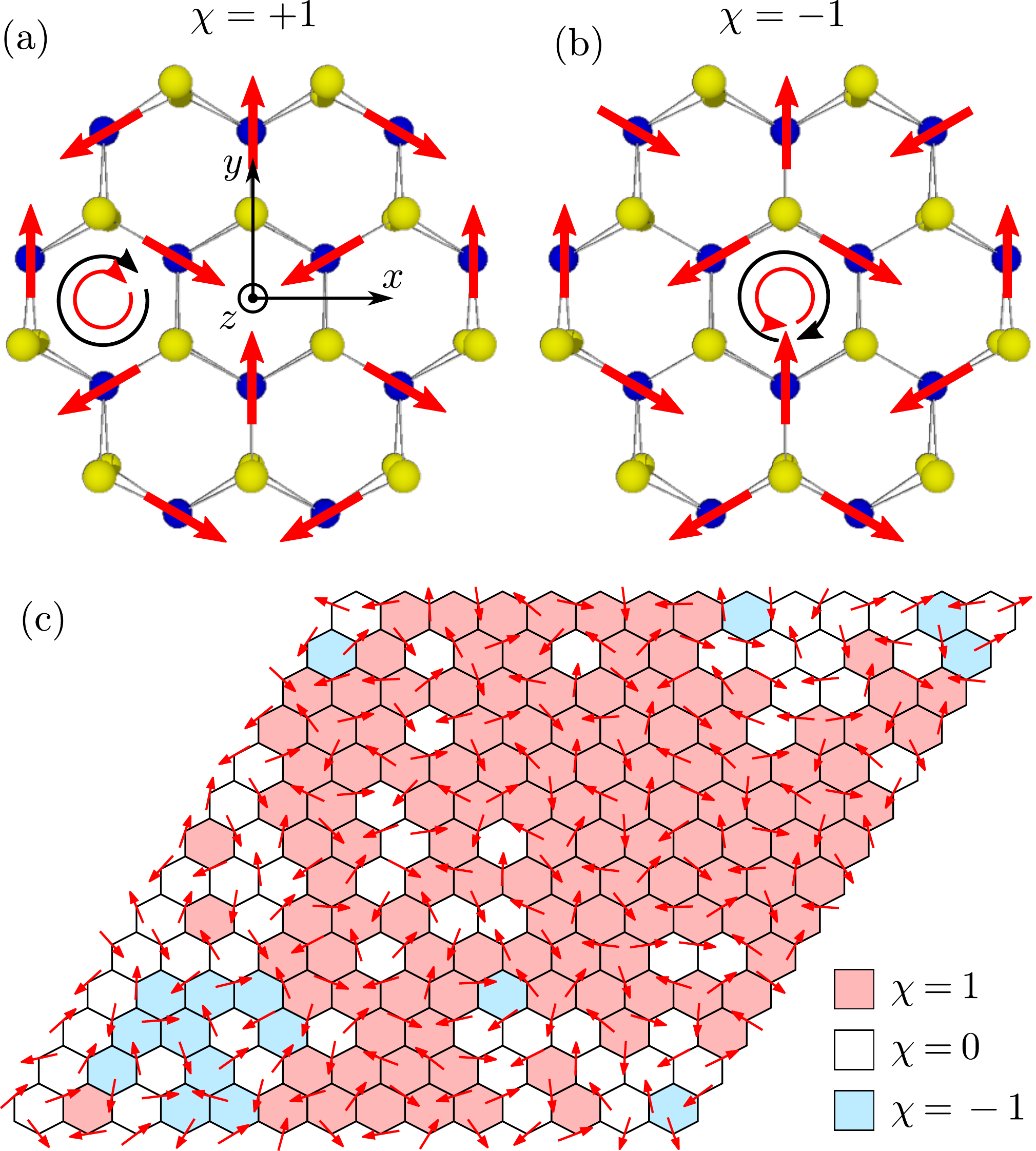}
  \caption{(a) Chiral $120^{\circ}$ antiferromagnetic N\'eel state with positive chirality $\chi=+1$. (b) The same state with negative chirality $\chi=-1$. (c) A snapshot taken from a Monte Carlo simulation of magnetic thermodynamics: the red, white and blue hexagons have chiralities $\chi_n=+1$, $0$, and $-1$, respectively.}
\label{fig:helix}
\end{figure}

\textit{Magnetic ground state and phase transitions.---}Now we move from pairwise interactions to the collective ground state of many impurities on top of a TMD ML. We assume that adatoms are located above every transition metal atom, as shown in Fig.~\ref{fig:helix}(a,b). This means, in fact, formation of a hybrid bilayer structure~\cite{macneill_NaturePhysics2017,shi_NatureNanotechnology2019,Ciorciaro2020,zhong_NatNanotechnol2020}, see discussion below. Then the RKKY interaction coexists with an additional ``direct'' exchange interaction. We describe the latter by the Hamiltonian
\begin{equation}
  \mathcal H_\text{dir}=\sum_{\langle i,j\rangle}\left[J_\perp S_{i,z}S_{j,z}+J_\parallel\left(S_{i,x}S_{j,x}+S_{i,y}S_{j,y}\right)\right],
\end{equation}
where the sum runs over the pairs of the nearest spins only, while $J_\perp$ and $J_\parallel$ are the strengths of the direct Ising and XY (in-plane) exchange interactions. The total Hamiltonian is given by $\mathcal H=\mathcal H_\text{ind}+\mathcal H_\text{dir}$.

The magnetic phase diagram of this system is generally complex. We focus on an interesting case of easy plane ($|V_\perp|\ll|V_\parallel|$, $|J_\perp|\ll J_\parallel$) and consider antiferromagnetic direct exchange, $J_\parallel>0$~\cite{Fang2021}. Then the spin thermodynamics is described by the planar rotator (XY) model, where the spins $\bm S_i$ are treated as classical in-plane vectors with their directions parameterized by angles $\theta_i$. This representation gives the total energy of the system in the form
\begin{equation}
  \label{eq:E}
  E=J\sum_{\langle i,j\rangle}\cos(\theta_i-\theta_j)+\sum_{i<j}V_s(R_{i,j})\cos(\theta_j-\theta_i-\varphi_{i,j}),
\end{equation}
where $J=J_\parallel S^2$ and $V_s(R)=V_\parallel S^2[J_0(k_FR)Y_0(k_FR)+J_1(k_FR)Y_1(k_FR)]$ is the scalar part of the in-plane RKKY interaction. The ground state for both terms is a chiral $120^{\circ}$ antiferromagnetic N\'eel state, which is shown in Fig.~\ref{fig:helix}(a).

The model defined by the first term in Eq.~\eqref{eq:E} has already been studied extensively. Its ground state was found to be two-parameter degenerate~\cite{Miyashita1984,Lee1984}: First, all the rotators can be simultaneously rotated by the same angle $\phi$, which is a continuous parameter. The Mermin-Wagner-Berezinskii theorem~\cite{Mermin1966,Berezinskii1971} forbids macroscopic breaking of this symmetry at any finite temperature, so the system experiences the Berezinskii-Kosterlitz-Thouless transition~\cite{_2012,kosterlitz_RepProgPhys2016} at temperature $T_s$, when the exponential decay of the correlations in $\theta(\bm r)$ becomes power-law.

\begin{figure*}
  \includegraphics[width=0.48\linewidth]{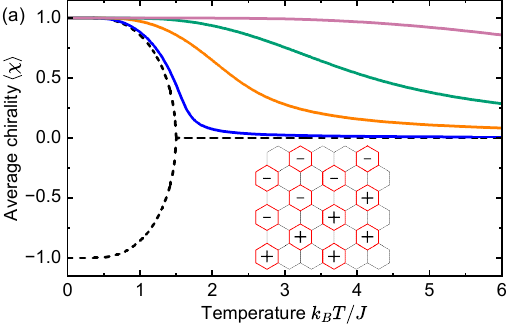}
  \hfill
  \includegraphics[width=0.48\linewidth]{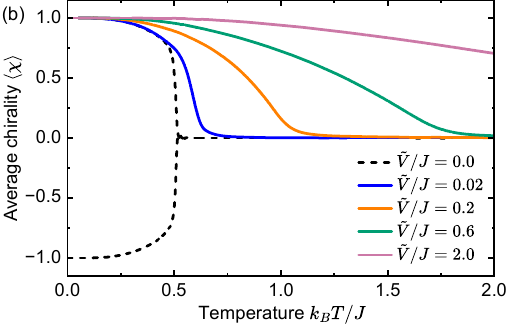}
  \caption{\label{fig:MFA}
    Temperature dependence of the average chirality for the different strengths of the RKKY interaction specified in the legend in (b). The black dashed lines show the two equivalent chiral phases in the absence of the RKKY interaction, $\tilde V=0$. (a)~Mean field approach of Eqs.~\eqref{eq:MFA} and~\eqref{eq:Delta_E}. The inset shows a domain wall between different chiralities used for the derivation of the mean field theory. (b) Monte Carlo simulations for $k_F a_0 = 0.1$ (and $L^2=27^2$ atoms).
  }
\end{figure*}

The second symmetry is discrete and corresponds to the two possible chiralities of the ground state, as shown in Fig.~\ref{fig:helix}(a,b). For each unit cell of the TMD ML, when one makes a round trip (black arrow), the magnetic moments on this path can rotate clockwise, counterclockwise, or not rotate on average. This corresponds to the chirality of the given hexagon $\chi_n$ equal to $+1$, $-1$, and $0$, respectively [see Fig.~\ref{fig:helix}(c)]. Due to the discrete nature of this order parameter, spontaneous breaking of the symmetry in $\chi$ gives rise to a second order (Ising-like) phase transition at temperature $T_c>T_s$~\cite{Obuchi2012}. The temperatures of the two phase transitions are very close to each other, $T_c\approx T_s\approx J/(2k_B)$, and differ approximately by 1\% only. The origin of this proximity is not completely clear~\cite{PhysRevLett.88.167007}.

The RKKY interaction changes this picture of the phase transitions qualitatively. First, we focus on the chirality-related phase transition, which can be analyzed in the mean field approach. To do this, we divide the lattice into non-overlapping hexagons, as shown in the inset in Fig.~\ref{fig:MFA}(a). The chirality-related phase transition is driven by the formation of domain walls between regions with the opposite chiralities~\cite{Lee1986,PhysRevLett.88.167007}, and its optimal geometry is shown in the inset in Fig.~\ref{fig:MFA}(a). In the End Matter Appendix \hyperref[app:MFA]{B} we show that the energy cost of the domain wall formation and the RKKY interaction energy between chiral hexagons ($\chi_n=\pm1$) are given by
\begin{equation}
  \label{eq:E_chi}
  E=-\frac{J}{4}\sum_{\braket{n,m}}\chi_n\chi_m+\frac{9}{4}\sum_{n<m}V_s(R_{n,m})(1+\chi_n)(1+\chi_m).
\end{equation}
The first term here describes Ising model on the triangular lattice which is very well studied~\cite{Houtappel1950,Wannier1950,Fisher1967}. The second term describes the long-range RKKY interaction and notably breaks the global symmetry $\chi_n\to-\chi_n$. Indeed, from Eq.~\eqref{eq:E} one can see directly that the RKKY interaction energy is non-zero only between the hexagons with the positive chirality, because it specifically favors clockwise spin rotations in each hexagon, see Fig.~\ref{fig:helix}(c).

At finite temperature, the average chirality can be calculated self-consistently in the mean field approach as
\begin{equation}
  \label{eq:MFA}
  \braket{\chi}=\tanh\frac{\Delta E}{2k_BT},
\end{equation}
where $\Delta E$ is the energy difference between $\chi_n=+1$ and $\chi_n=-1$ states, assuming that all other hexagons have the same average chirality $\braket{\chi}$. Explicitly, Eq.~\eqref{eq:E_chi} gives
\begin{equation}
  \label{eq:Delta_E}
  \Delta E=3J\braket{\chi}+2\sqrt{3}\tilde{V}(1+\braket{\chi}),
\end{equation}
where $\tilde V=V_\parallel S^2/(k_Fa_0)^2$ and we took into account that $k_Fa_0\ll 1$, so the summation in the second term in Eq.~\eqref{eq:E_chi} can be replaced with an integration. This expression reveals two important aspects: (i) The long-range nature of the RKKY interaction leads to a large factor $(k_Fa_0)^{-2}$ in $\tilde V$, so even though the interaction between two neighbouring spins is relatively weak ($V_s(a_0)\sim V_\parallel\ln[1/(k_Fa_0)]$), the total RKKY contribution to the interaction energy in the ordered phase can be substantial. (ii) The chiral nature of the RKKY interaction makes the states with positive and negative chirality inequivalent. In particular, the energies of the completely chiral states shown in Fig.~\ref{fig:helix}(a,b) differ by $2\tilde{V}/\sqrt{3}$ per unit cell.

The result of the calculation of the average chirality after Eqs.~\eqref{eq:MFA} and~\eqref{eq:Delta_E} is shown in Fig.~\ref{fig:MFA}(a). The black dashed lines show that in the absence of the RKKY interaction, there is a second order phase transition at $T_c=3J/(2k_B)$. Above this temperature the average chirality vanishes and below $T_c$ it takes either positive or negative value. In the limit of zero temperature, one has $\braket{\chi}=\pm1$. But even very weak RKKY interaction breaks equivalence between the two chiral states and favors $\braket{\chi}>0$ at any finite temperature, as shown by the blue curve. 

A better quantitative understanding of the role of the RKKY interaction can be reached by means of Monte Carlo simulations of finite clusters (the simulation code is publicly available \cite{MX2RKKYjl}). The details of modeling are given in App.~\ref{app:MonteCarlo} and the result of the average chirality calculation is shown in Fig.~\ref{fig:MFA}(b) (for $k_Fa_0=0.1$, which corresponds to $E_F\approx 8.5$~meV with $a_0=3$~{\AA} and an electron effective mass equal to half the free electron mass). One can see a good qualitative agreement with the results of the mean field model. In particular, in the limit of $\tilde V=0$, there is a second order phase transition, which is correctly predicted by the mean field theory. The main difference is that the critical temperature is $T_c\approx J/(2k_B)$~\cite{Miyashita1984,Lee1984,Obuchi2012}, which is smaller than in the mean field approach as usual~\cite{White2007}. The nonzero chirality at all temperatures at $\tilde V>0$ means formation of a smooth crossover instead of the phase transition due to the RKKY interaction.

\begin{figure}
  \includegraphics[width=\linewidth]{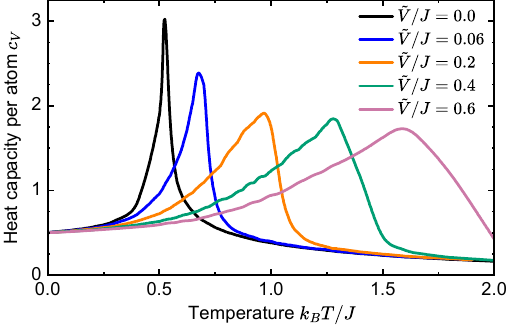}
  \caption{\label{fig:MC}
    Heat capacity per atom calculated within the Monte Carlo simulations for different values of the RKKY interaction for the same parameters as in Fig.~\ref{fig:MFA}(b).
  }
\end{figure}

Importantly, the Monte Carlo simulations allow us to study also the effect of the RKKY interaction on the Berezinskii–Kosterlitz–Thouless spin orientation phase transition. The RKKY interaction does not break the continuous rotational symmetry of the spin system, therefore the long-range ordering remains forbidden by the Mermin--Wagner theorem at any finite temperature. The phase transition can be evidenced from the temperature dependence of the specific heat, as shown in Fig.~\ref{fig:MC}. In the absence of the RKKY interaction, there is a single peak at approximately the same temperature as for the chiral phase transition~\cite{Miyashita1984,Lee1984,Obuchi2012}. With increase of the RKKY interaction strength, the peak shifts to higher temperatures, broadens and somewhat decreases in amplitude. In the End Matter we show that the heat capacity per atom saturates with increase of the sample size and that the fluctuations of the absolute value of magnetization are the largest approximately at the same temperature, where the heat capacity has a peak. All this points to the shift of the Berezinskii–Kosterlitz–Thouless phase transition to higher temperatures. This happens because the RKKY interaction favors the same 120$^\circ$ N\'eel state, enhances the long-range correlations, but does not break the rotational symmetry.

\textit{Discussion.---}Magnetism in TMD MLs has already attracted significant attention in both experimental and theoretical studies across a variety of settings~\cite{zhang_AdvancedScience2020,pham_AdvancedMaterials2020,wang_Nature2022}. In particular, the RKKY interaction in these systems has been investigated previously~\cite{Parhizgar2013,Hatami2014,Mastrogiuseppe2014,Losada2023}. However, earlier works either neglected spin-valley locking or relied upon simplified and not realistic choices of atomic orbitals. Consequently, the chiral RKKY interaction Hamiltonian specific to electrons in the lower conduction subband~\eqref{eq:V_tensor} was missing.

Interestingly, it has been suggested that the spin orientation related phase transition in the absence of the RKKY interaction is non-universal~\cite{Obuchi2012}. In contrast, when direct exchange interactions are neglected in Eq.~\eqref{eq:E}, the planar rotator model can be mapped, via a local transformation, onto the conventional ferromagnetic XY model with an effective interaction range on the order of $1/k_F$~\footnote{More precisely, the spin stiffness is very small and negative on a discrete lattice, but a small direct exchange interaction makes it positive.}. This model is expected to exhibit a universal Berezinskii-Kosterlitz-Thouless phase transition. Hence, by tuning the strength of the RKKY interaction, one can explore a crossover from non-universal to universal critical behavior---an unusual and intriguing scenario that calls for further investigation beyond the scope of the present work.

The proposed chiral magnetic ordering is quite general. On the one hand, it might be realized in hybrid van-der-Waals heterobilayers, specifically with VCl$_2$, VI$_2$, NiGa$_2$S$_4$, CrSe$_2$, and Hf$_2$VC$_2$F$_2$ which demonstrate 120$^\circ$ antiferromagnetic N\'eel order, have lattice constants close to those of TMD MLs, possess wide band gaps preventing charge transfer, can be realized in the ML limit~\cite{Liu_PRL2024,Li_NatMater2021,Victorin_SciRep2024,Zhang_JACS2018}. At the same time, at present it is not possible to indicate a single perfect candidate; thus, further DFT and experimental studies are needed to establish the feasibility of such a structure. The typical phase transitions temperatures in these materials are tens of Kelvin and may be slightly increased by the RKKY interaction.

On the other hand, the same physics can be realized for an ensemble of magnetic molecules deposited on a TMD ML~\cite{Fang2021}. In the latter case, the coverage may be incomplete, which would reduce the exchange interaction, but not change the chiral exchange interaction qualitatively, as we demonstrate in Supplemental Material.

The emergence of a chiral $120^{\circ}$ N\'eel magnetic state strongly influences the optical and transport properties of a TMD ML. From Fig.~\ref{fig:helix}(a) one can see that the chiral magnetic order enlarges the magnetic unit cell, leading to folding of the Brillouin zone for electrons. This, in turn, enables coupling between electronic states in opposite valleys. In Ref.~\cite{i.s.krivenko_jointsubmission2026a} we describe experimental signatures of such coupling: exciton fine structure splitting and brightening of intervalley excitons. In transport, the chiral magnetic texture manifests through an antiferromagnetic anomalous Hall effect, whose magnitude can be controlled via the chirality and orientation of the magnetization. These phenomena can be controlled by tuning the electron Fermi level.

As an outlook, we note that the quantum spin liquid state~\cite{savary_RepProgPhys2016,zhou_RevModPhys2017,broholm_Science2020} was initially proposed by Anderson for the spin $1/2$ on a triangular lattice~\cite{anderson_MaterialsResearchBulletin1973}. The later studies demonstrated that the actual state is almost classical~\cite{bernu_PhysRevLett1992,PhysRevLett.82.3899,white_PhysRevLett2007}, which supports the validity of our classical description of the spin thermodynamics even for magnetic moments as small as $1/2$. However, the quantum spin liquid emerges if the next nearest neighbor interactions are taken into account~\cite{kaneko_JPhysSocJpn2014,hu_PhysRevB2015,zhu_PhysRevB2015}. Although it was recognized that the scalar chiral interactions can stabilize quantum spin liquids~\cite{hu_PhysRevB2016,wietek_PhysRevB2017,drescher_2025}, we anticipate that the chiral RKKY interaction will suppress the quantum spin liquid state and favor the $120^{\circ}$ N\'eel magnetic order. Deeper investigations are needed to check this conjecture.

\textit{In conclusion,} we have demonstrated that spin-valley locking in TMD MLs gives rise to a chiral RKKY interaction between magnetic adatoms with the strength tunable by the electron Fermi energy. This interaction favors 120$^\circ$ N\'eel order and introduces a new model to study phase transitions in atomically thin magnets. When combined with the nearest neighbor antiferromagnetic coupling, the chiral RKKY interaction lifts the degeneracy between states with opposite chiralities. Using the mean field theory and Monte Carlo simulations, we have shown that this leads to the suppression of the Ising-like chiral phase transition and to the stabilization of a unique chirality even for weak indirect interaction. In contrast, the spin orientation related Berezinskii–Kosterlitz–Thouless(-like) phase transition persists but shifts to higher temperatures. These results establish valley-controlled exchange as a powerful mechanism for engineering chiral magnetism, with promising implications for van der Waals magnetic heterostructures.


\let\oldaddcontentsline\addcontentsline
\renewcommand{\addcontentsline}[3]{}

\textit{Acknowledgements.---}We thank A.V. Shumilin for fruitful discussions. I.S.K. acknowledges funding from the European Research Council (ERC) under the European Union's Horizon 2020 research and innovation programme (Grant agreement No.\ 854843-FASTCORR).
He also gratefully acknowledges the computing time made available to him on the high-performance computer ``Lise'' at the NHR center NHR@ZIB (Starter application 25226). This center is jointly supported by the Federal Ministry of Research, Technology, and Space and the state governments participating in the NHR (\url{www.nhr-verein.de}).
V.N.M. thanks Russian Science Foundation for support via grant No. 25-12-00093.
The development of the mean field theory by D.S.S. was supported by the Russian Science Foundation Grant No. 25-72-10031.

\renewcommand{\i}{\ifr}
%

\renewcommand{\i}{{\rm i}}


\newpage
\onecolumngrid
\begin{center}
\large\bfseries End Matter
\end{center}
\twocolumngrid

\appendix

\refstepcounter{section}
\textit{Appendix A: Derivation of RKKY interaction.---}The bare Green's function of electrons in the lower subband of the TMD ML conduction band has the form
\begin{equation}
  \label{RKKYDerivation}
  G_{s,s'}^{(0)}(\omega,\bm r)=\delta_{s,s'}\int\frac{\e^{\i\left(\bm{k}+s\bm K\right)r}}{\hbar\omega-\frac{\hbar^2k^2}{2m}+E_F+\i 0\sign(\omega)}\frac{\d\bm k}{(2\pi)^2}.
\end{equation}
It differs from the standard Green's function of 2D electrons only by a factor $\e^{\i s\bm{Kr}}$, which reflects the spin-valley locking. The integration gives
\begin{multline}
  \label{eq:G_ss1}
  G_{s,s'}^{(0)}(\omega,\bm r)=-\frac{m\delta_{s,s'}}{\pi\hbar^2}\exp(\i s\bm{Kr})\\
  \times K_0\left(-\i\sign(\omega)r\sqrt{2m(E_F+\hbar\omega)}/\hbar\right).
\end{multline}

The static spin susceptibility of the electron gas can be expressed as
\begin{equation}
  \chi_0^{\alpha,\beta}(\bm r)=\i\hbar\int\frac{\d\omega}{2\pi}\sigma_{s,s'}^\alpha G_{s',s''}^{(0)}(\omega,\bm r)\sigma_{s'',s'''}^\beta G_{s''',s}^{(0)}(\omega,-\bm r).
\end{equation}
With the notation $G_{s}^{(0)}(\omega,\bm r)=G_{s,s}^{(0)}(\omega,\bm r)$, this coincides with Eq.~\eqref{eq:chi_main}. Substituting here Eq.~\eqref{eq:G_ss1} we obtain
\begin{equation}
  \chi_0^{\alpha,\beta}(\bm r)=-\sigma_{s,s'}^\alpha\e^{\i s'\bm{Kr}}\sigma_{s',s}^\beta\e^{-\i s\bm{Kr}}\frac{\tilde{\chi}_0(r)}{2},
\end{equation}
with the standard static density (instead of spin) susceptibility of a 2D electron gas~\cite{Fischer1975,BalMonod1987,Litvinov1998}
\begin{multline}
  \label{eq:chi_dens}
  \tilde{\chi}_0(r)=-\i\hbar\int\frac{\d\omega}{2\pi}G_{s,s}^{(0)}(\omega,\bm r)G_{s,s}^{(0)}(\omega,-\bm r)\\=\frac{mk_F^2}{2\pi\hbar^2}\left[J_0(k_Fr)Y_0(k_Fr)+J_1(k_Fr)Y_1(k_Fr)\right],
\end{multline}
(here summation over $s$ is assumed).

Finally, we substitute this susceptibility into Eq.~\eqref{eq:A2} and make use of the following exchange interaction tensor
\begin{equation}
  \label{eq:As}
  A=\begin{pmatrix}
    A_\parallel & 0 & 0 \\
    0 & A_\parallel & 0 \\
    0 & 0 & A_\perp
  \end{pmatrix},
\end{equation}
which takes into account the symmetry of the electron orbitals at the metal atoms~\cite{Avdeev2019}. This brings us to Eqs.~\eqref{eq:H_RKKY} and~\eqref{eq:V_tensor}.


\refstepcounter{section}
\label{app:MFA}
\textit{Appendix B: Mean field theory.---}To derive the mean field theory, we divide the whole lattice into non-overlapping hexagons, as shown by red in the inset in Fig.~\ref{fig:MFA}(a). In the vicinity of the phase transition, we assume that each of these hexagons has a chirality $\chi_n=\pm1$.

First, let us analyze the energy of direct exchange interaction neglecting RKKY interaction and reduce it to the energy of interaction between the nearest hexagons. In the ground state, all hexagons have the same chirality. The chirality related phase transition is driven by formation of domain walls between regions with the opposite chirality of hexagons. The optimal geometry of the domain wall was found in Ref.~\onlinecite{Lee1986}, and it is shown in the inset in Fig.~\ref{fig:MFA}(a). One can see that each hexagon at this wall interacts with two hexagons on the opposite side of the wall with the opposite chirality. The loss of energy per period in this configuration is $J$ as compared to the ground state~\cite{Lee1986}. Therefore, the interaction energy between hexagons of the opposite chirality is by $J/2$ larger than between hexagons of the same chirality. This gives the following expression for the direct exchange interaction energy between chiral hexagons,
\begin{equation}
  \label{eq:E_J}
  E_J=-\frac{J}{4}\sum_{\braket{n,m}}\chi_n\chi_m,
\end{equation}
where the sum runs over all pairs of neighboring hexagons forming the triangular lattice. 
The critical temperature in this model is known exactly and equals $T_c=J/(k_B\ln 3)$~\cite{Houtappel1950,Wannier1950,Fisher1967}.

\begin{figure*}
  \includegraphics[width=0.48\linewidth]{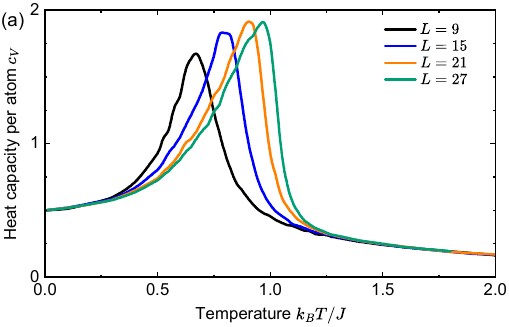}
  \hfill
  \includegraphics[width=0.48\linewidth]{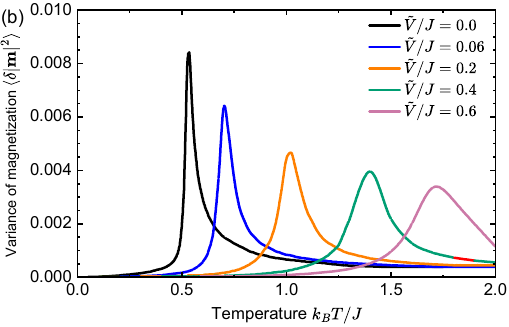}
  \caption{\label{fig:convergence}
    (a) Heat capacity per atom calculated within the Monte Carlo simulations for different cluster sizes specified in the legend for $\tilde V/J=0.2$ and $k_Fa_0=0.1$. (b) Variance of the magnetization magnitude $\braket{\delta|\bm m|^2}$ calculated for the same parameters as in Fig.~\ref{fig:MC}.
  }
\end{figure*}

Second, we consider the RKKY interaction between the hexagons. We assume that two spins belonging to the same hexagon form an angle $2\pi/3$ with each other. The RKKY interaction~\eqref{eq:V_tensor} favors positive chirality with exactly the $2\pi/3$ angle between nearest spins. By contrast, this interaction energy vanishes (for $k_Fa_0\ll 1$) when a given spin interacts with a triad of spins in a hexagon with negative chirality. As a result, only hexagons with positive chirality interact with each other by this mechanism. Since there are three spins in each hexagon, the RKKY interaction energy takes the form
\begin{equation}
  \label{eq:E_V}
  E_V=\frac{9}{4}\sum_{n<m}V_s(R_{n,m})(1+\chi_n)(1+\chi_m),
\end{equation}
where $R_{n,m}$ denotes the distances in the pairs. This expression neglects the RKKY interaction inside the hexagons, which is small as compared with the long range interaction between many spins in different hexagons. Note that $V_s(R)$ diverges only logarithmically at small $R$.

The sum of Eqs.~\eqref{eq:E_J} and~\eqref{eq:E_V} gives Eq.~\eqref{eq:E_chi}.

\refstepcounter{section}
\label{app:MonteCarlo}

\textit{Appendix C: Details of Monte Carlo simulations.---}The Monte Carlo simulations are performed for the model~\eqref{eq:E} defined on a finite parallelogram-shaped cluster of the triangular lattice with linear sizes of $L\times L$ atoms, subject to periodic boundary conditions. In all simulations, $L$ was taken to be a multiple of 3, as the corresponding cluster would not be able to support the $120^{\circ}$ antiferromagnetic N\'eel order otherwise.

Following the study of Obuchi and Kawamura~\cite{Obuchi2012}, we employed the standard classical Monte Carlo method with the Metropolis single-spin update combined with the over-relaxation update. The simulations were organized as a series of $N_{MC}\sim10^3$ steps, each comprising a complete sequential Metropolis sweep of the cluster followed by a complete over-relaxation sweep and a measurement of averages ($\braket{\chi}_{MC}$, $\braket{E}_{MC}$, etc.). Measurement results from the first 10\% of the steps were discarded for the purpose of equilibration. Temperature scans were performed starting from the lowest end of the scanned ranges, where the initial spin configuration was manually set to the $120^{\circ}$ N\'eel state with $\chi=+1$. Simulations for each subsequent point of the $T$-range used the final spin configuration from the previous $T$-point as their starting configuration.

A special remark should be made about the treatment of the RKKY interaction~\eqref{eq:V_tensor}, whose range is formally infinite. In our simulations, the range was effectively truncated to $L/2$. A pair of cluster spins $i$ and $j$ interacts via the RKKY term $\propto V_s(R_{i,j})$, where the distance $R_{i,j}$ is defined as the shortest distance between $i$ and $j$ computed while taking into account the periodic boundary conditions. Geometrically speaking, $R_{i,j}$ is the length of the shortest straight path on the surface of a torus. This approach is justified if the interaction is mostly localized within the cluster, $L \gtrsim 1/(k_F a_0)$. Besides, our definition of the distance $R_{i,j}$ guarantees that in the fully ordered state all spins contribute the same amount to the RKKY energy regardless of their position in the cluster (boundary vs interior spins).


The heat capacity per spin was calculated using the unbiased estimate of the variance of the total energy $E$,
\begin{equation}
  c_V(T)=\frac{\langle(\delta E)^2\rangle}{L^2T^2} =
  \frac{N_{MC}}{N_{MC}-1} \frac{\braket{E^2}_{MC} - \braket{E}_{MC}^2}{L^2 T^2}.
\end{equation}
It is shown in Fig.~\ref{fig:convergence}(a) as a function of the cluster size. One can see that the heat capacity saturates with increase of the cluster size, which is consistent with the Berezinskii--Kosterlitz--Thouless phase transition. Additional calculation results are shown in Supplemental Material.

Another observable of interest measured over the course of the Monte Carlo simulations was the variation of the magnetization magnitude $|\bm m|$ at the quasimomentum point $K_+$,
\begin{equation}
  \label{eq:m_var}
    \braket{\delta|\bm m|^2} =
    \frac{N_{MC}}{N_{MC}-1}
    \left(\braket{|\bm m|^2}_{MC} - \braket{|\bm m|}_{MC}^2\right),
\end{equation}
where the magnetization is defined as
\begin{equation}
    \bm m = \frac{1}{L^2} \sum_i \bm S_i e^{-i \bm K_+ \bm R_i}.
\end{equation}

The temperature dependence of $\braket{\delta|\bm m|^2}$ is shown in Fig.~\ref{fig:convergence}(b) for different strengths of the RKKY interaction. One can see that it has a sharp peak that shifts together with the peak in the heat capacity in Fig.~\ref{fig:MC}. In addition we note that due to the direct exchange interaction, the spin stiffness is positive, and at finite temperature the range of the RKKY interaction is quenched. All this supports our claim that the RKKY interaction does not destroy the Berezinskii–Kosterlitz–Thouless(-like) spin phase transition, but shifts it to higher temperatures.

\let\addcontentsline\oldaddcontentsline
\makeatletter
\renewcommand\tableofcontents{%
    \@starttoc{toc}%
}
\makeatother
\renewcommand{\i}{{\rm i}}

\onecolumngrid
\vspace{\columnsep}
\begin{center}
\newpage
\makeatletter
{\large\bf{Supplemental Material to\\``\@title''}}
\makeatother
\end{center}
\vspace{\columnsep}



\counterwithin{figure}{section}
\renewcommand{\section}[1]{\oldsec{#1}}
\renewcommand{\thepage}{S\arabic{page}}
\renewcommand{\theequation}{S\arabic{equation}}
\renewcommand{\thefigure}{S\arabic{figure}}
\renewcommand{\bibnumfmt}[1]{[S#1]}
\renewcommand{\citenumfont}[1]{S#1}

\setcounter{page}{1}
\setcounter{section}{0}
\setcounter{equation}{0}
\setcounter{figure}{0}

\begin{figure*}[b]
  \includegraphics[width=0.48\linewidth]{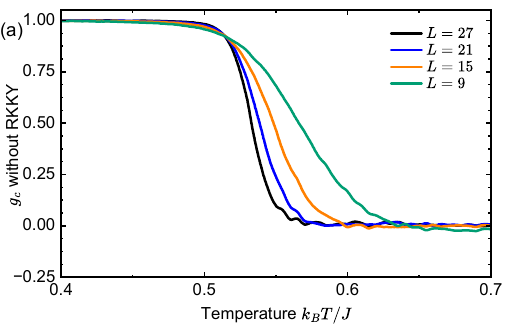}
  \hfill
  \includegraphics[width=0.48\linewidth]{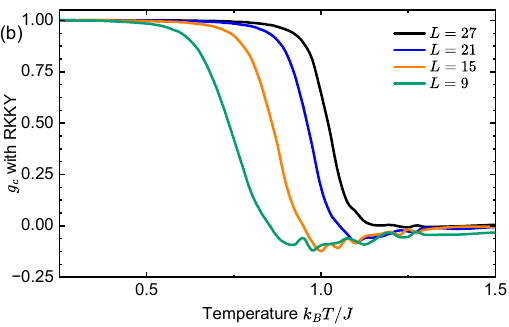}
  \includegraphics[width=0.48\linewidth]{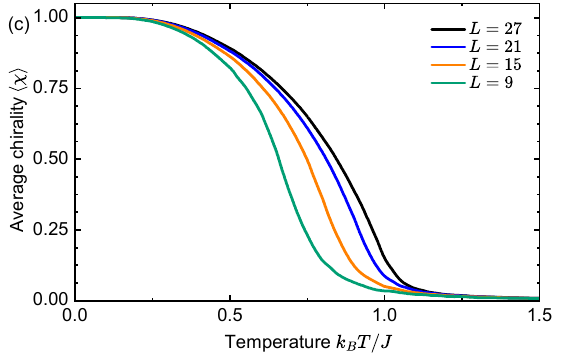}
  \hfill
  \includegraphics[width=0.48\linewidth]{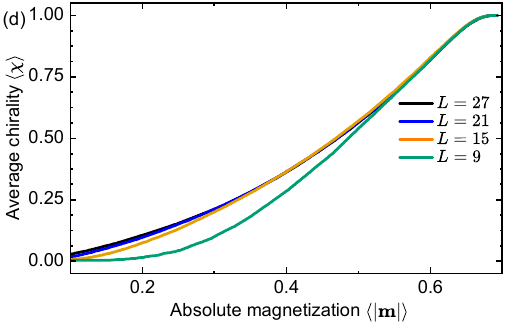}
  \caption{\label{fig:quad}
    (a) Chiral binder parameter, Eq.~\eqref{eq:gc} as a function of temperate for the different cluster sizes specified in the legend without RKKY interaction, $\tilde V=0$. (b) The same as in (a), but for $\tilde V=0.2J$. (c)~Temperature dependence of the average chirality for the same parameters as in (b). (d) Average chirality as a function of the average absolute magnetization $\braket{\left|\bm m\right|}$ for the same parameters as in (b).
  }
\end{figure*}

From Fig.~\ref{fig:convergence}(a) one can see that the maximum of the heat capacity converges already for $L=27$ despite its moderate ratio with the effective RKKY interaction length $1/(k_Fa_0)\sim10$. At the same time, there is a slow shift of the temperature, where heat capacity reaches its maximum. To study this in more detail, we show in Fig.~\ref{fig:quad} additional results of the Monte Carlo simulations.

Fig.~\ref{fig:quad}(a) and (b) show the chiral Binder parameter
  \begin{equation}
    \label{eq:gc}
    g_c=\frac{3}{2}\left(1-\frac{\left\langle X^4\right\rangle}{3\left\langle X^2\right\rangle^2}\right)
  \end{equation}
  with the total chirality
  \begin{equation}
    X=\sum_n\chi_n
  \end{equation}
  (the sum runs over all $L^2$ hexagons) for the cases of $\tilde V=0$ and $\tilde V=0.2J$. In the former case $g_c$ for different $L$ cross near $k_BT=J/2$ for all $L$, which evidences second order chiral phase transition at this temperature. By contrast, at $\tilde V>0$, this common crossing disappears due to lifting of chiral ($Z_2$) symmetry by the RKKY interaction.

Still one can see the slow convergence of the temperature dependence of the average chirality in Fig.~\ref{fig:quad}(c). This convergence shows that larger $L$ may be useful to obtain the ultimate limiting curve, however, it will be close to the case of $L=27$ without any qualitative differences.

To have a deeper insight, we plot in Fig.~\ref{fig:quad}(d) the average chirality against the average magnetization $\braket{\left|\bm m\right|}$ at $K_+$ point. One can see that the curves for $L=15$, $21$, and $27$ almost perfectly overlap. This evidences that the finite size of the cluster effectively renormalizes the strength of the interaction without changing the relation between different magnetic parameters.

\begin{figure*}
  \includegraphics[width=0.48\linewidth]{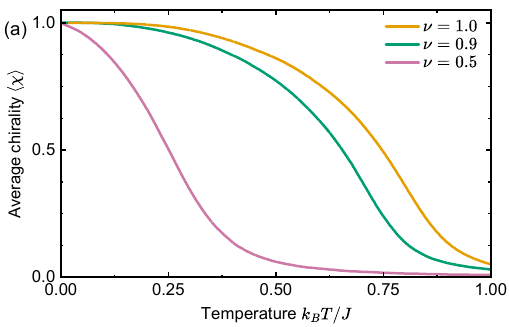}
  \hfill
  \includegraphics[width=0.48\linewidth]{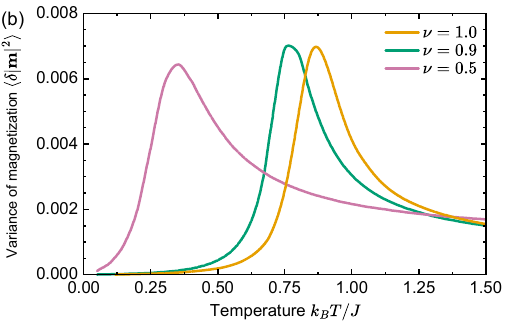}
  \caption{\label{fig:covering}
    (a) Average chirality and (b) variance of the magnetization absolute value as functions of the temperatures for the different fillings of the sites by magnetic impurities $\nu$, specified in the legend. Averaging over 100 disorder realizations is performed after calculation of variance for each realization.
  }
\end{figure*}

Further, we extend the numerical model to include vacancies, which is relevant for the magnetic adatoms on top of a TMD ML. In Fig.~\ref{fig:covering} we compare the previous calculations with the cases of $\nu=90$\% and $50\%$ of occupied sites by magnetic impurities. One can see that both temperature dependences of the average chirality and of the variance of the magnetization absolute value [Eq.~\eqref{eq:m_var} with normalization by the number of impurities] shift to lower temperatures with decrease of $\nu$, because both direct exchange and RKKY interactions get suppressed. Note that we exclude the hexagons with incomplete filling from the calculation of the average chirality.

\begin{figure*}
  \includegraphics[width=0.48\linewidth]{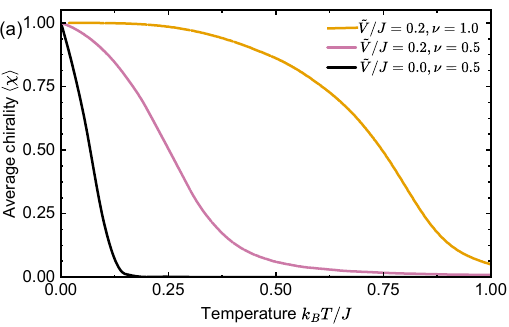}
  \hfill
  \includegraphics[width=0.48\linewidth]{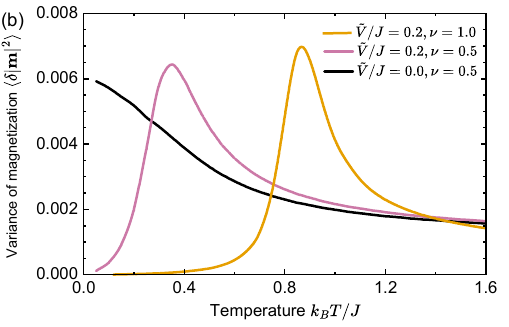}
  \caption{\label{fig:wo_RKKY}
    The same as in Fig.~\ref{fig:covering} for $\nu=1$ and $\nu=0.5$ with additional black curves corresponding to the calculations for $\nu=0.5$ without RKKY interaction with averaging over 10 disorder realizations.
  }
\end{figure*}

In order to separate the effects of two different exchange interactions, we show in Fig.~\ref{fig:wo_RKKY} by the black curves the results of an additional calculation for $\nu=0.5$ but without RKKY interaction. This case is very specific, because isolated clusters of magnetic impurities form in it that do not interact with each other. Effectively, the system becomes an ensemble of the finite size systems. As a result, the average chirality very quickly decreases with increase of the temperature, Fig.~\ref{fig:wo_RKKY}(a). And more importantly, the fluctuations of magnetization absolute value $|\bm m|$ do not vanish at zero temperature being the largest, see black curve in Fig.~\ref{fig:wo_RKKY}(b). This evidences the absence of the BKT transition in ensemble of finite clusters without RKKY interaction. However, even a small long range RKKY interaction couples different clusters and recovers the BKT transitions, as shown by the purple line in Fig~\ref{fig:wo_RKKY}(b).

\end{document}